\documentclass[twocolumn, 10pt, journal]{IEEEtran}

\IEEEoverridecommandlockouts
\usepackage{lineno}
\usepackage[font=footnotesize]{caption}  
\usepackage{subcaption}

\usepackage{amsmath}
\usepackage{cite}
\usepackage{float}
\usepackage{lettrine}
\usepackage{soul}
\usepackage{color}
\usepackage[table]{xcolor}
\usepackage{url}
\usepackage{diagbox}
\usepackage{multirow}
\usepackage{array}
\usepackage{bm}
\usepackage{physics}
\usepackage[overload]{empheq}
\usepackage{comment}

\usepackage{amsmath}
\usepackage{graphicx}
\usepackage{booktabs} 
\usepackage{times}
\usepackage{fancyhdr,amsmath,amssymb}

\usepackage{amsmath, amssymb}
\usepackage{makecell}
\usepackage{multirow}
\usepackage{enumitem}
\usepackage{xcolor}
\usepackage{soul}
\sethlcolor{yellow}

\newcolumntype{M}[1]{>{\centering\arraybackslash}m{#1}}
\newcolumntype{N}[1]{>{\raggedright\arraybackslash}m{#1}}
\newcommand{\eat}[1]

\usepackage{tikz}

\usepackage[usestackEOL]{stackengine}

\stackMath

\usepackage{booktabs}

\usepackage{enumitem}

\usepackage{amsmath}
\usepackage{graphicx}
\usepackage{booktabs} 
\usepackage{times}
\usepackage{fancyhdr,amsmath,amssymb}

\usepackage{amsmath, amssymb}
\usepackage{makecell}
\usepackage{multirow}
\usepackage{enumitem}
\usepackage{xcolor}
\usepackage{soul}
\sethlcolor{yellow}
\usepackage{tikz}

\usepackage[usestackEOL]{stackengine}
\stackMath
\usepackage{overpic}

\usepackage{booktabs}

\usepackage{enumitem}
\usepackage{threeparttable}
\usepackage{tabularx}
\newcommand{\PreserveBackslash}[1]{\let\temp=\\#1\let\\=\temp}
\newcolumntype{C}[1]{>{\PreserveBackslash\centering}p{#1}}
\newcolumntype{R}[1]{>{\PreserveBackslash\raggedleft}p{#1}}
\newcolumntype{L}[1]{>{\PreserveBackslash\raggedright}p{#1}}

\graphicspath{ {fig/} }

\usepackage{algpseudocode}

\begin{document}

\title{DAE-Embedded Neural Control Verification for Shipboard Microgrids under Transient Shocks}

\author{ Fei~Feng,  Lizhi~Wang and Ziqian Liu
\thanks{ Fei~Feng and Ziqian Liu are with the Department of Electrical Engineering, State University of New York, Maritime College, Bronx, NY 10465, USA (Corresponding author e-mail: ffeng@sunymaritime.edu).}
\thanks{Lizhi~Wang is with Siemens Foundational Technologies, Princeton, NJ 08540, USA.}}

\markboth{}
{Shell \MakeLowercase{\textit{ et al.}}:  Bare Demo of IEEEtran.cls for IEEE Journals}

\maketitle

\begin{abstract} 
Neural control offers strong potential for handling highly nonlinear dynamics in shipboard microgrids (SMGs), yet its black-box nature can trigger abrupt control spikes and actuator saturation during initial transient shocks. This letter devises a formal verification method for SMG neural controller to assess its shock responses. Our contributions include: 1) a set-based SMG differential–algebraic equation(DAE) model compatible with set propagation; 2)  a DAE-embedded bound propagation approach to  compute tight envelopes of all possible neural control output. Extensive case studies demonstrate the effectiveness of the devised method in formally certifying SMG control performance under uncertain disturbances.
\end{abstract}

\IEEEpeerreviewmaketitle

\section{Introduction}
\IEEEPARstart{M}{aritime} transportation accounted for nearly 90\% of the value of global overseas trade over the past decade\cite{Skjong2016}. The 2024 collapse of the Francis Scott Key Bridge in Baltimore, caused by the containership \textit{MV Dali} losing propulsion power, highlights the urgent need to modernize ship energy systems \cite{MillerHooks2024}. Neural control has recently emerged as a promising paradigm for handling highly nonlinear dynamics to enable safer shipboard microgrid (SMG) operation.

However, neural control still risks abrupt control spikes and actuator saturation under sudden transient shocks because: 1) the inherent generalization gap between finite training data and the infinite state-disturbance space encountered in real SMG operations; 2) highly nonlinear black-box nature can output extreme value at the onset of large disturbances. Unfortunately, neither the existing direct methods nor time-domain simulation methods can effectively address the impact of those uncertainties on ship power systems\cite{Kundur1994}. Probabilistic time-domain simulations such as Monte Carlo sampling are prohibitively expensive in considering uncertainties from complex marine conditions. 

Recognizing that the most critical safety risk of black-box neural control lies in its instantaneous overreaction at the   onset of extreme disturbances, we focus on verifying this initial transient phase. To this end, this letter devises a DAE-embedded safety verification method (DBBP) for SMG neural control that encloses all possible control responses triggered by sudden shocks in a single analysis.



\section{DAE-embedded Verification Framework of SMG Neural  Controls} 
Propulsion predominates over 70\% of ship power demands and varies significantly under uncertain marine conditions. This letter considers a typical SMG with one propulsion unit and two synchronous generators (SGs) equipped with automatic voltage regulators (AVRs) and neural controllers for voltage and frequency stabilization. The proposed verification framework is generic and applicable to other SMG models. \vspace{-10pt} 

\subsection{Set-based SMG-DAE Formulation}\label{sec:IIA_model}

\textbf{1) Set-based hydro-propulsion model:}
To ensure compatibility with bound propagation, a set-based induction motor (IM) model is formulated from IM dynamics \cite{10716430},
\begin{equation}\label{eq:e_IM}
\begin{bmatrix}
\mathbf{e}_{dq}^{k+1} \\
\omega_{im}^{k+1}
\end{bmatrix}
=
\begin{bmatrix}
(1 - c_1)\mathbf{e}_{dq}^k 
- c_1(x - x')\mathbf{J}\mathbf{i}_{dq}^k 
- c_2 s^k \mathbf{J}\mathbf{e}_{dq}^k \\[6pt]
(1 - c_3 k_{f,im})\omega_{im}^k 
+ c_3 \left( T_e^k - \mathcal{T}_L \right)
\end{bmatrix}
\end{equation}
where $c_1=\Delta t/T_0'$, $c_2=\Delta t\omega_b$, and $c_3=\Delta t/(2H_{im})$; 
$\mathbf{J}=\left[\begin{smallmatrix}0&-1\\1&0\end{smallmatrix}\right]$ is the rotation matrix; 
$\mathbf{e}_{dq}^k$ and $\mathbf{i}_{dq}^k$ denote the transient internal voltage and stator current sets in the $dq$ frame; 
$x,x'$ are reactances, $T_0'$ is the transient time constant, and $s^k$ is the rotor slip; 
$T_e^k$ is electromagnetic torque, $\omega_{im}^k$ is shaft speed, $H_{im}$ is inertia, and $k_{f,im}$ is friction. 
The propulsion load is bounded by $\mathcal{T}_L=T_{L0}(\omega_{im})\oplus\Delta\mathcal{T}_L$, 
where $\Delta\mathcal{T}_L(\kappa_w,R_T,V_s)$ captures marine uncertainties $\Delta\mathcal{T}_L$ such as wake fraction $\kappa_w$, hull resistance $R_T$, and propeller torque coefficient $K_Q$ \cite{10716430}.

\textbf{2) Set-based neural-controlled SG model:}
SGs are regulated by neural controllers and AVRs, with the neural output superimposed on the AVR error loop to determine the excitation input. The SG dynamics\cite{10716430} are flattened into white-box computational graph,
\begin{equation}\label{eq:SG}
\begin{bmatrix}
\delta^{k+1} \\
\omega^{k+1} \\
e_q'^{k+1}
\end{bmatrix}
=
\begin{bmatrix}
\delta^k + c_4 (\omega^k - 1) \\
(1 - c_5 D)\omega^k + c_5 ( P_m - \mathcal{P}_e^k ) \\
(1 - c_6)e_q'^k + c_6 \left[ \mathcal{E}_{fd}^k - (x_d - x_d')\mathcal{I}_d^k \right]
\end{bmatrix}
\end{equation}
where $c_4 = \Delta t \omega_b$, $c_5 = \Delta t / (2H)$, and $c_6 = \Delta t / T_{d0}'$, $H$ is the inertia, $D$ is the damping coefficient, and $T_{d0}'$ is the $d$-axis transient time constant. $\mathcal{P}_e^k$ and $\mathcal{I}_d^k$ are the bounded sets of electromagnetic power and $d$-axis current, respectively, derived from the real-valued power flow algebraic equations.

To bridge the physical dynamics with the cyber control layer, we aggregate the aforementioned differential state set (IM and SG dynamics) and algebraic state set (power flows and voltages)  within $\mathcal{X}^k$ and $\mathcal{Y}^k$, respectively. Given the bounded observation $\mathcal{O}^k \subseteq \{\mathcal{X}^k, \mathcal{Y}^k\}$, a deep neural controller $\mathcal{U}^k =\pi_\theta(\mathcal{O}^k)$ is embeded into AVR to provide intelligent stabilization against severe marine dynamics.

A major bottleneck in formulating set-based AVR model lies in the non-differentiable hard clipping imposed by the AVR output saturation block $[0, E_{max}]$ (i.e., the exciter ceiling and floor limiters). We transform the hardware saturation into a dual-ReLU activation structure. Given the terminal voltage set $\mathcal{V}_t^k$ and the neural controller's pre-clamp output set $\mathcal{U}^k$, the excitation demand is $\mathcal{E}_{dem}^k = K_A (V_{ref} - \mathcal{V}_t^k + \mathcal{U}^k)$. The actual saturated excitation set $\mathcal{E}_{fd}^k$ injected into \eqref{eq:SG} is formulated as,
\begin{equation}\label{eq:AVR}
    \mathcal{E}_{fd}^k = \text{ReLU}(\mathcal{E}_{dem}^k) - \text{ReLU}(\mathcal{E}_{dem}^k - E_{max})
\end{equation}

This devised ReLU-based relaxation enables to directly penetrate the physical actuator limits without gradient truncation, thereby evaluating whether the neural controller reaches saturation under transient shocks.

\textbf{3) Set-based SMG-DAE model:}  SGs and propulsion unit are electrically coupled by combining \eqref{eq:e_IM}-\eqref{eq:AVR} with the SMG admittance network, yielding the set-based SMG-DAE model
\begin{equation}\label{eq:DAE_diff}
    \mathcal{X}^{k+1} = \mathbf{A}\mathcal{X}^k \oplus \mathbf{B}\Phi_{bi}^k \oplus \mathbf{C}\Phi_{re}^k \oplus \mathbf{E}\mathcal{W}^k
\end{equation}
\begin{equation}\label{eq:DAE_alg}
    \mathcal{Y}^k = \mathbf{M}\mathcal{X}^k \oplus \mathbf{D}\mathcal{W}^k
\end{equation}
where $\mathcal{X}^k=[\delta^k;\omega^k;e_q'^k;\mathbf{e}_{dq}^k;\omega_{im}^k]$ and $\mathcal{Y}^k=[\mathcal{V}_t^k;\mathcal{I}_d^k]$ denote the differential and algebraic state sets, respectively, and $\mathcal{W}^k$ represents hydro uncertainties (e.g., $\Delta\mathcal{T}_L$). 
$\mathbf{A},\mathbf{B},\mathbf{C},\mathbf{E}$ are the state transition, bilinear coupling, actuator mapping, and disturbance matrices, while $\mathbf{M}$ and $\mathbf{D}$ are derived from the inverted network admittance. 
$\Phi_{bi}^k\!\triangleq\!\Phi_{bi}(\mathcal{X}^k,\mathcal{Y}^k)$ captures bilinear spatial couplings (e.g., slip–voltage products in IM dynamics), and $\Phi_{re}^k\!\triangleq\!\Phi_{re}(\mathcal{U}^k,\mathcal{Y}^k)$ encodes actuator saturation via the dual-ReLU representation in \eqref{eq:AVR}.


\subsection{DAE-Embedded Backward Bound Propagation}

Forward reachability often suffers from severe wrapping effects when propagating through coupled neural network and SMG nonlinearities. To mitigate this, the set-based DAE model is embedded into neural network as a dummy neural layer to form a unified verification model, enabling backward bound propagation via linear relaxations across the closed-loop system.

\textbf{1) DAE–neural unified verification model:}
The SMG dynamics \eqref{eq:DAE_alg} are integrated into the neural observation as,
\begin{equation}
\mathcal{U}^k
=
\pi_\theta(\mathcal{O}^k)
=
\pi_\theta \!\left(
\mathbf{H}_x \mathcal{X}^k
\oplus
\mathbf{H}_y (\mathbf{M}\mathcal{X}^k \oplus \mathbf{D}\mathcal{W}^k)
\right)
\end{equation}
where $\pi_\theta$ denotes the neural controller and $\mathbf{H}_x,\mathbf{H}_y$ are linear extraction matrices. 
The control set is injected into the DAE update \eqref{eq:DAE_diff} via the dual-ReLU saturation operator $\Phi_{re}^k(\mathcal{U}^k,\mathcal{Y}^k)$, forming the closed-loop verification graph.

\textbf{2) Backward bound propagation:}
Rather than simulating trajectories, we propagate bounds backward from the target control output $\mathcal Z_{out}\in \mathcal{U}^k$ to the disturbance $\mathcal{W}^k$. For any intermediate signal $\mathbf z$ in the DAE–neural graph, we assume an enclosure with respect to $\mathcal W^k$,
\begin{equation}
\underline{\mathbf{\gamma}}^\top \mathcal W^k + \underline b
\;\le\;
\mathbf z
\;\le\;
\overline{\mathbf{\gamma}}^\top \mathcal W^k + \overline b
\end{equation}
where $\underline{\mathbf{\gamma}},\overline{\mathbf{\gamma}}$ are disturbance sensitivity vectors and $\underline b,\overline b$ are bias terms. The following primed quantities (e.g., $\mathbf z', \underline{\mathbf{\gamma}}',\overline{\mathbf{\gamma}}',\underline b',\overline b'$) denote updated coefficients after propagation.

\emph{(a) Nonlinear relaxation.}
For any nonlinearity $\mathbf z'=\sigma(\mathbf z)$ in $\pi_\theta$ or $\Phi_{re}^k$, we construct linear relaxations\cite{zhang2019stableefficienttrainingverifiably} as,
\begin{equation}\label{eq:nonlinearz}
\underline{\mathbf{\Lambda}} \mathbf z + \underline{\mathbf{\Delta}}
\le
\sigma(\mathbf z)
\le
\overline{\mathbf{\Lambda}} \mathbf z + \overline{\mathbf{\Delta}}
\end{equation}
where $\underline{\mathbf{\Lambda}},\overline{\mathbf{\Lambda}}$ are diagonal slope matrices and $\underline{\mathbf{\Delta}},\overline{\mathbf{\Delta}}$ are intercept vectors. 
Substituting the enclosure of $\mathbf z$ into \eqref{eq:nonlinearz} yields the updated disturbance sensitivities,
\begin{equation}
\underline{\mathbf{\gamma}}' =
\underline{\mathbf{\Lambda}}^+ \underline{\mathbf{\gamma}}
+
\underline{\mathbf{\Lambda}}^- \overline{\mathbf{\gamma}},
\quad
\overline{\mathbf{\gamma}}' =
\overline{\mathbf{\Lambda}}^+ \overline{\mathbf{\gamma}}
+
\overline{\mathbf{\Lambda}}^- \underline{\mathbf{\gamma}}
\end{equation}
with bias terms updated by $\underline{\mathbf{\Delta}},\overline{\mathbf{\Delta}}$; $\underline{\mathbf{\Lambda}}^+ = \max(\underline{\mathbf{\Lambda}},0),
\underline{\mathbf{\Lambda}}^- = \min(\underline{\mathbf{\Lambda}},0)$, and $\overline{\mathbf{\Lambda}}^+$ and $\overline{\mathbf{\Lambda}}^-$ share the same rules.

\emph{(b) Linear transformation.}  For any linear mapping $\mathbf z' = \mathbf L \mathbf z + \mathbf c$, where $\mathbf{L}$ represents either the neural network weights or the physical system matrices (e.g., $\mathbf{L} \in \{\mathbf{M}, \mathbf{A}, \mathbf{B}, \mathbf{C}, \mathbf{D}, \mathbf{E}\}$), and $\mathbf{c}$ denotes the constant offsets in the neural layer biases and the steady-state operating point vectors in the DAE-neural model,
\begin{equation}
\underline{\mathbf{\gamma}}' =
\mathbf L^+ \underline{\mathbf{\gamma}}
+
\mathbf L^- \overline{\mathbf{\gamma}},
\quad
\overline{\mathbf{\gamma}}' =
\mathbf L^+ \overline{\mathbf{\gamma}}
+
\mathbf L^- \underline{\mathbf{\gamma}}
\end{equation}
\begin{equation}
\underline b' =
\mathbf L^+ \underline b
+
\mathbf L^- \overline b
+
\mathbf c,
\quad
\overline b' =
\mathbf L^+ \overline b
+
\mathbf L^- \underline b
+
\mathbf c
\end{equation}
where $\mathbf L^+ = \max(\mathbf L,0)$ and $\mathbf L^- = \min(\mathbf L,0)$ are the elementwise positive and negative parts.

By recursively applying rules (a)–(b) from the neural output through the neural network and DAE updates, every signal admits an enclosure in $\mathcal W^k$. The accumulated coefficients yield the global disturbance-to-control relation,
\begin{equation}
\underline{\gamma}^\top\mathcal W^k+\underline b
\le \mathcal Z_{out} \le
\overline{\gamma}^\top\mathcal W^k+\overline b
\end{equation}

This closed-form formulation  isolates the maximum possible control deviation under severe marine conditions, proving the saturation-free property of the neural controller without relying on heuristic sampling.

\begin{figure}[t]
  \centering
  \includegraphics[width=0.49\textwidth]{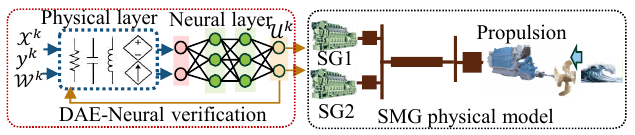}
  \caption{SMG neural control verification architecture} 
  \label{LR_architecture1}
\end{figure}
\begin{figure}[t]
\centering
\begin{minipage}{0.98\columnwidth}
\vspace{0.4em}\hrule\vspace{0.4em}
\textbf{Algorithm 1:} \text{DBBP Algorithm}\\[-0.6em]
\hrule\vspace{0.4em}
\begin{algorithmic}[1]
\State \textbf{Input:} DAE–neural SMG model,  $\mathcal W^k$, target $\mathcal Z_{out}$
\State Initialize $(\underline{\gamma},\overline{\gamma},\underline b,\overline b)$ for $\mathbf z $, set Eqs. (4)-(6)
\For{layers $\ell=L,\dots,1$ (backward)}
  \If{$\ell$ nonlinear}
    \State Update $(\underline{\gamma},\overline{\gamma},\underline b,\overline b)$ via relaxation slopes/intercepts
  \Else
    \State Update coefficients via affine map $\mathbf L$
  \EndIf
\EndFor
\State \textbf{Output:} $\underline{\gamma}^\top\mathcal W^k+\underline b
\le \mathcal Z_{out} \le
\overline{\gamma}^\top\mathcal W^k+\overline b$
\end{algorithmic}
\vspace{0.3em}\hrule
\end{minipage}
\end{figure}

\begin{table}[t]
\centering
\caption{SMG System Parameters}
\label{tab:system_parameters}
\footnotesize
\begin{tabular}{ccccccc}
\toprule
\textbf{Para.} & \textbf{SG1} & \textbf{SG2} & \textbf{Para.} & \textbf{AVR} & \textbf{Para.} & \textbf{IM} \\
\midrule
$H$ (s)            & 1.5 & 3.0  & $K_A$              & 20  & $H_{im}$ (s)      & 0.8 \\
$D$                & 2.0 & 1.5  & $V_{ref}$              & 1.0  & $X_s$ (p.u.)      & 0.1 \\
$x_d'$ (p.u.)      & 0.4 & 0.55 & $E_{\max}$ (p.u.)  & 7   & $X_m$ (p.u.)      & 3.0 \\
$T'_{d0}$ (s)      & 5.0 & 6.0  & $\Delta t$ (s)     & 0.05& $X_r$ (p.u.) & 0.1 \\
\bottomrule
\end{tabular}
\end{table}

\begin{table}[t]
\centering
\caption{Empirical vs. Verified Control Bounds}
\label{tab:bounds_comparison}
\begin{tabular}{clccc}
\toprule
\textbf{Control} & \textbf{Signal} & \textbf{MC Empirical Bounds} & \textbf{DBBP Bounds}  \\
\midrule
\multirow{2}{*}{\shortstack{\textbf{NN1}\\stable }} 
& SG1  & $[-0.09496, -0.09494]$ & $[-0.09496, -0.09481]$  \\
& SG2 & $[0.09484, 0.09485]$ & $[0.09365, 0.09487]$  \\
\midrule
\multirow{2}{*}{\shortstack{\textbf{NN2}\\ overreact }} 
& SG1  & $[0.01431, 0.04578]$ & $[0.00971, 0.04710]$  \\
& SG2  & $[0.05101, 0.06228]$ & $[0.04578, 0.06300]$  \\
\bottomrule
\end{tabular}
\end{table}

\section{Case Studies}
 The effectiveness of the devised DBBP method is validated on a typical SMG with 2 SGs and 1 propulsion unit (see Fig. \ref{LR_architecture1}). The SGs are equipped with pre-trained deep neural controllers to enhance transient stability, while the external hydrodynamic shocks are injected through the propulsion induction motor to emulate abrupt marine disturbances.  Key SMG parameters are listed in Table.\ref{tab:system_parameters}. The neural controller $\pi_\theta$ is implemented as a multi-layer perceptron (MLP) mapping a 7-dimensional observation vector $\mathcal{O}^k$ to a 2-dimensional supplementary excitation action bounded in $[-0.1, 0.1]$ p.u. 

\subsection{Validity of DBBP }

This subsection validates the effectiveness of the devised verification framework. Table.\ref{tab:bounds_comparison} compares the output bounds of two neural controllers (NN1: stable, NN2: overreacting) obtained from DBBP verification and Monte Carlo (MC) simulations. A sudden propulsion load shock is emulated by a $+20\%$ nominal load step ($\Delta T_m = 0.1$ p.u.) with uncertainty $\epsilon = \pm 0.02$ p.u. The following insights are observed:

\begin{itemize}[leftmargin=*] 
\item  { In both cases, the verification bounds enclose the MC sampling results, validating the correctness of the proposed method. For example, in the overreacting case, the DAE-based verification yields a higher upper bound and a lower lower bound for SG1 compared with MC sampling.  } 
\item  { The bounds obtained by the proposed method closely envelopes the MC bounds, demonstrating its tightness. For instance, in the stable case, the maximum errors for SG1 and SG2 are 0.000128 and 0.001198, respectively; in the overreacting case, the errors are 0.0046 and 0.0052.} 
\end{itemize}

 \begin{figure}[t]
  \centering
  \includegraphics[width=0.48\textwidth]{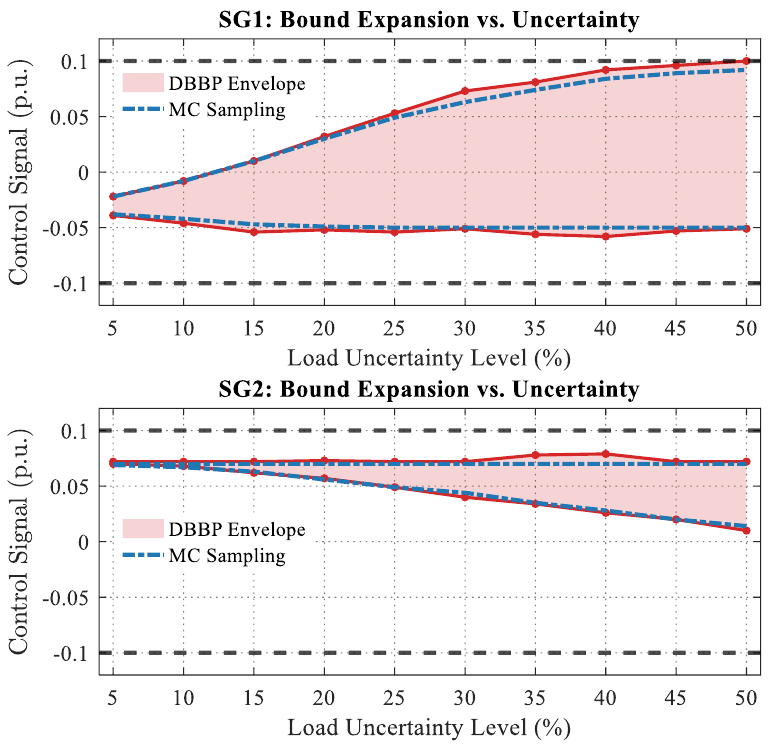}
  \caption{Bound verification under different uncertainty levels} 
  \label{LR_architecture}
\end{figure}
\subsection{DBBP Results under Different Uncertainty Levels}
This subsection is to evaluate the scalability of the proposed framework under different hydrodynamic uncertainties. Fig. \ref{LR_architecture} compares the bound expansion of NN2  obtained by DBBP with that from MC sampling. The hydrodynamic load uncertainty is increased from  $5\%$ to $50\%$. Some insights can be found:
\begin{itemize}[leftmargin=*] 
    \item Across different uncertainty levels, the DBBP envelope strictly encloses the empirical MC extremes, validating its scalability and assessing that worst-case control spikes are close to limitations under severe perturbations.
    \item The DBBP bounds for both SG1 and SG2 remain within the actuator saturation limits of $\pm 0.1$ p.u. (black dashed lines) across all load uncertainty levels. This provides a rigorous safety certificate that the neural controller maintains stability without triggering hardware saturation under both nominal and extreme maritime disturbances.
\end{itemize}

\section{Conclusion}

This letter proposes a DAE-embedded backward bound propagation framework to formally certify neural control safety in shipboard microgrids under transient shocks. By integrating a set-based DAE-neural computational model with bound propagation, the method provides tight enclosures of neural control actions, explicitly accounting for physical couplings and actuator saturations. Case studies demonstrate accurate and scalable disturbance-to-control certification under hydrodynamic uncertainties. Future work will extend the framework to SMG neural network design.

\bibliographystyle{IEEEtran}
\bibliography{ref}

\end{document}